\documentclass[conference]{IEEEtran}
\IEEEoverridecommandlockouts
\makeatletter
\newcommand{\linebreakand}{%
  \end{@IEEEauthorhalign}
  \hfill\mbox{}\par
  \mbox{}\hfill\begin{@IEEEauthorhalign}
}
\makeatother
\usepackage{booktabs}
\usepackage{cite}
\usepackage{amsmath,amssymb,amsfonts}
\usepackage{algorithmic}
\usepackage{graphicx}
\usepackage[ruled, linesnumbered, boxed]{algorithm2e} % For algorithms

\usepackage{textcomp}
\usepackage{xcolor}
\def\BibTeX{{\rm B\kern-.05em{\sc i\kern-.025em b}\kern-.08em
    T\kern-.1667em\lower.7ex\hbox{E}\kern-.125emX}}
\begin{document}

\title{Automating Physics-Based Reasoning for SysML Model Validation\\
% Automated Validation of SysML Diagrams Generated from Physics-Based Rules
}

\author{\IEEEauthorblockN{Candice Chambers}
\IEEEauthorblockA{\textit{Department of Computer Science} \\
\textit{Florida Institute of Technology}\\
Melbourne, USA \\
chambersc2017@my.fit.edu}
\and
\IEEEauthorblockN{Summer Mueller}
\IEEEauthorblockA{\textit{Department of Computer Science} \\
\textit{Florida Institute of Technology}\\
Melbourne, USA \\
smueller2023@my.fit.edu }
\and
\IEEEauthorblockN{Parth Ganeriwala}
\IEEEauthorblockA{\textit{Department of Computer Science} \\
\textit{Florida Institute of Technology}\\
Melbourne, USA \\
pganeriwala2022@my.fit.edu}
\linebreakand
\IEEEauthorblockN{Chiradeep Sen}
\IEEEauthorblockA{\textit{Department of Mechanical Engineering} \\
\textit{Florida Institute of Technology}\\
Melbourne, USA \\
csen@fit.edu}
\and
\IEEEauthorblockN{Siddhartha Bhattacharyya}
\IEEEauthorblockA{\textit{Department of Computer Science} \\
\textit{Florida Institute of Technology}\\
Melbourne, USA \\
sbhattacharyya@fit.edu}
}

\maketitle

\begin{abstract}

System and software design benefits greatly from formal modeling, allowing for automated analysis and verification early in the design phase. Current methods excel at checking information flow and component interactions, ensuring consistency, and identifying dependencies within Systems Modeling Language (SysML) models. However, these approaches often lack the capability to perform physics-based reasoning about a system's behavior represented in SysML models, particularly in the electromechanical domain. This significant gap critically hinders the ability to automatically and effectively verify the correctness and consistency of the model's behavior against well-established underlying physical principles. Therefore, this paper presents an approach that leverages existing research on function representation, including formal languages, graphical representations, and reasoning algorithms, and integrates them with physics-based verification techniques. Four case studies (coffeemaker, vacuum cleaner, hairdryer, and wired speaker) are inspected to illustrate the model's practicality and effectiveness in performing physics-based reasoning on systems modeled in SysML. This automated physics-based reasoning is broken into two main categories: (i) structural, which is performed on BDD and IBD, and (ii) functional, which is then performed on activity diagrams. This work advances the field of automated reasoning by providing a framework for verifying structural and functional correctness and consistency with physical laws within SysML models. 

\end{abstract}


\section{Introduction}

Model-Based Systems Engineering (MBSE) with SysML is gaining traction for designing complex systems \cite{kirshner2022integrating}. MBSE emphasizes the need for developing analytical models early in the design phase to enable automated reasoning. This has led to the creation of structured and formal method-based approaches. For example, safety-critical industries such as aerospace and defense companies have been designing and developing model-based approaches with models developed in Architecture Analysis and Design Language (AADL) \cite{stewart2017, chauhan2022toward} and SysML \cite{BhattacharyyaDASC14, chambers2023towards,chambers2022natural} to provide artifacts generated from the design phase itself. Similarly, model-based approaches have been developed for the modeling and analysis of specifications in cybersecurity \cite{alfageehUEMCOM2019} \cite{gupta2023modeling}. Architecture modeling and analysis enables graphical and formal representation of system components, interactions, and workflows, allowing engineers to evaluate design consistency, functional dependencies, model verification and validation early in development \cite{zhang2021towards}. However, ensuring these models' validity, especially regarding fundamental physical laws, remains challenging \cite{ganeriwala2023functional}.

Traditional SysML validation techniques primarily address logical and structural aspects, like component interactions and information flow consistency \cite{hecht2021verification}. While effective for identifying flaws in component dependencies or data flows, these methods often fall short in enforcing physics-based constraints \cite{zhang2023system}. Consequently, SysML models may appear correct in terms of connectivity and structure but still deviate from real-world physical principles—such as conservation of energy or balance laws of transport—which are critical in domains like electromechanical systems.

To address this limitation, an automated approach was proposed to embed physics-based reasoning within SysML model validation, bridging the gap between functional model correctness and adherence to physical laws. Our method integrates formal function representation, graphical analysis, and reasoning algorithms, enabling SysML models to be validated not only for logical consistency but also for physical validity. This approach extends formalism in functional representations from Ganeriwala et al. \cite{ganeriwala2023functional}. Specifically, our approach categorizes automated reasoning into two main aspects: (i) structural reasoning, applied to SysML’s Block Definition Diagrams (BDD) and Internal Block Diagrams (IBD), and (ii) functional reasoning, applied to Activity Diagrams representing dynamic workflows. By automating physics-based reasoning in this way, the framework facilitates comprehensive verification of a system’s structural and functional alignment with physical laws.

Four case studies were examined across various domains, including electromechanical and energy systems, to demonstrate the practicality and effectiveness of this approach. These case studies illustrate how automated physics-based reasoning can detect inconsistencies that might be missed by conventional SysML validation techniques, enhancing model reliability and reducing the risk of costly redesigns later in the development process. In section \ref{sec:rw} related work is discussed, with discussion on the representation of structure and functions in systems modeling in section \ref{sec:sfrep}. In section \ref{sec:physReasoning}, the automated reasoner defines and implements various inspections. The algorithmic design of automated functions reasoner is explained in section \ref{sec:autoReason}. Then, in section \ref{sec:valid}, the automated reasoner performs the proposed validation on four use cases of electromechanical systems (coffeemaker, hair dryer, vacuum cleaner, and wired speaker) with the conclusion in section \ref{sec:conclusion}.

\section{Review of Related Works}
\label{sec:rw}
% Candice
Mahani et al. present an automatic approach for verifying finite state machines (FSM) modeled in SysML using the model checking tool New Symbolic Model Verifier (NuSMV) for a vehicular system \cite{mahani2021automatic}. NuSMV is a state-of-the-art model-checking tool developed at Carnegie Mellon University \cite{cimatti1999nusmv}. Model checking is a well-known formal verification technique that rigorously proves the correctness of a system based on its design requirements. This work is built using Cameo System Modeler (CSM) \cite{casse2017sysml}. Therefore, this work installed a Cameo plugin that transforms the SysML FSM into NuSMV and replays its verifications in the Cameo environment as an executable SysML State Machine. This greatly assists designs in extracting all verification errors in the FSM and mitigating them promptly. 

Conversely, Zhu et al. proposed an approach that uses MBSE to model an avionic system and then maps this formal framework to capture the functional requirements to support the ARP4754A guidelines \cite{zhu2019formal}. According to their proposed methodology, the black box operational scenario is modeled using SysML Sequence Diagrams; the functional scenario and chains are modeled using SysML Activity diagrams. Therefore, functional interface requirements can be captured from there. Ultimately, the top-level functional chains model is simulated to validate continuous functional chains against operational scenarios. 

Similarly, Zhang et al. proposed an approach that bridges the gap of traditional MBSE's inability to integrate domain-specific simulation tools \cite{zhang2021towards}. Domain-specific tools aid with removing ambiguities, inconsistencies, and challenges in the early design stage. Therefore, incorporating a descriptive system model with detailed domain-specific simulations allows for accurate design analysis. This approach lets designers validate the system's functional and non-functional requirements early in the design life cycle.

Yildirim et al. propose using the Enhanced Sequence Diagram (ESD) to improve the functional analysis of complex, multi-disciplinary systems \cite{YildirimRED2020}. This effort focused on interconnecting sequence diagrams from various use cases using parameters. The ESD framework is demonstrated using an electric hybrid bicycle case study. Through these scenarios, the ESD method proves effective for capturing and analyzing functional requirements, supporting rigorous functional modeling early in the design process, and enabling systems architecture development.

Mao et al. present an algorithmic approach to perform physics-based reasoning for the functional decomposition of engineered systems \cite{mao2024formal}. Functional decomposition is an imperative step in the early system engineering and design phase based on its components or subassembles functions. Typically, functional decomposition is done manually; however, this work proposed a three-part algorithm that has been validated to decompose black box models. In conjunction with topological reasoning, these three sections are used to synthesize and assemble the function structure graph.

Currently, there are no automated physics-based reasoners for validating functional and non-functional systems modeled in SysML. As seen in the works discussed above, most validation tools usually use reasoning on structural and functional diagrams based on the functional requirements set by stakeholders. Therefore, these systems are not being validated against physics-based laws but rather on the requirements placed during the early design stages. Therefore, this paper's main contribution is providing an automated reasoner that emphasizes performing reasoning on structural and functional models to adhere to physics-based rules. 

% \subsection{Research Gaps and Limitations}
% The research limitations of the previous work that are addressed in this effort are: 
% \begin{itemize}
%     \item Physics-based reasoning
%     \item Include
%     \item Include

% \end{itemize}

\section{Structural and Functional Representation in Systems Modeling}
\label{sec:sfrep}
% Parth
\subsection{System Modeling Language}
The System Modeling Language (SysML) is an extension of the Unified Modeling Language (UML) as it addresses UML's limitations while expanding its ability to analyze complex systems \cite{SysML}. SysML is also known as a dialect for UML 2 and defined as a UML 2 Profile. A UML Profile is a UML dialect that customizes the language via three mechanisms: Stereotypes, Tagged Values, and Constraints. SysML enables the specification, analysis, design, verification, and validation of various systems and systems-of-systems, an enabling technology for MBSE. System engineers typically use SysML to communicate various systems: software, hardware, information, processes, and personnel. While SysML removes the software-specific restrictions of UML, making it easier to learn, it still requires much time to construct these models. 

\subsection{Structural SysML Diagrams}
\subsubsection*{Block Definition Diagram }
Block Definition Diagram (BDD) \cite{enterprise} are graphical representations used in SysML that describe the structure and composition of a system. In a BDD, blocks represent the major components of a system along with their relationships and interactions. Blocks can be considered an abstract representation of the related functions or capabilities. Each block has a name, properties, ports, tag values depicting the block's function, and a set of relationships with other blocks. A block's properties describe the block's characteristics and sub-components instantiated blocks using internal block diagrams. 

\subsubsection*{Internal Block Diagram}
Internal Block Diagram (IBD) \cite{enterprise} describes the internal structure of a block element portrayed in the BDD. IBD demonstrates how blocks are connected and communicate within that block. In an IBD, blocks represent a system's major components, and the decomposition into smaller parts (subcomponents), called properties. Connectors are used to show the relationships between properties. The connectors show how data flows between the blocks or how power or signals are transmitted. These connectors can be directed or undirected, depending on the nature of the relationship between the properties. Relationships between blocks are represented in the IBD by association connectors, which are lines that connect the blocks.

\subsection{Functional SysML Diagrams}
\subsubsection*{Activity Diagram}
Traditionally, Activity Diagrams (ACT) \cite{enterprise} are used to represent functional representation in SysML. ACTs capture how different activities, tasks, or processes within a system are executed, showing both the control and data flow between them. Activity Diagrams are particularly useful for modeling workflows and processes where decisions, parallel activities, and synchronization points are necessary. In a SysML Activity Diagram, activities represent actions performed by system components, and these actions are connected via control flows that define the sequence of execution. Each activity can have input and output pins that act as buffers for the inputs (e.g., data or physical materials) needed by the activity and the outputs produced by the activity. These items can be physical entities such as energy, materials, or data, depending on the system under analysis.

\section{Physics-Informed Functional Reasoning}
\label{sec:physReasoning}
After representing a system's structural and functional components in SysML, these models are inspected against physics-based principles.
\subsection*{Inspection I: Balance Law}
\begin{figure} [h!tbp]
    \centering
    \includegraphics[width=\linewidth]{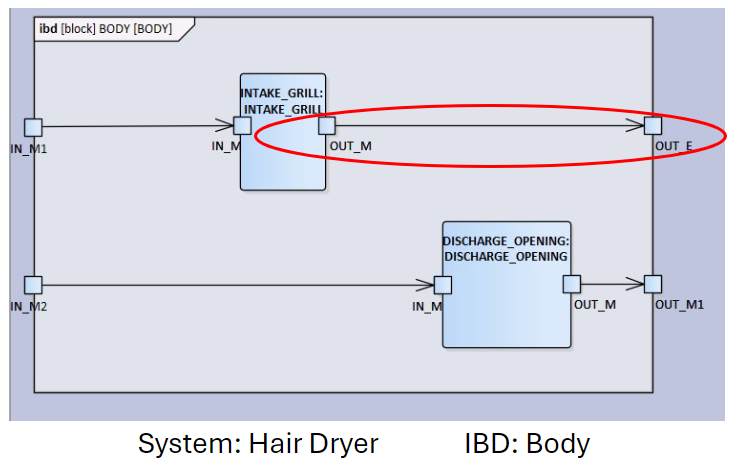}
    \caption{In the body IBD, the intake grill shows a material flowing (Out\_M) into input flow of type E (OUT\_E)}
    \label{fig:BL1-Hairdryer}
\end{figure}

\begin{figure} [h!tbp]
    \centering
    \includegraphics[width=\linewidth]{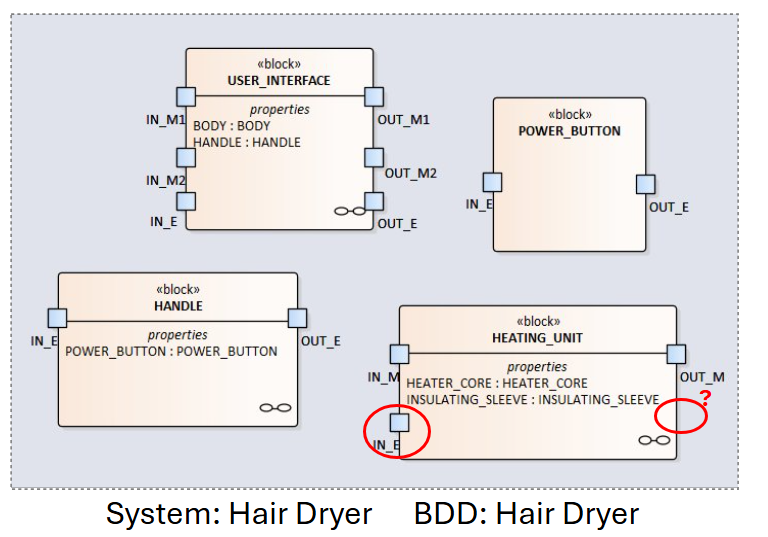}
    \caption{In the Hair Dryer block there is a barren flow as there is no output flow for type E (Out\_E)}
    \label{fig:BL2-Hairdryer}
\end{figure}

Inspection I, curated from Sen et al. \cite{sen2013physics}, specifically ensures that models maintain the balance laws of transport phenomena. This physics-based reasoning, directly and indirectly, examines the model's agreement with the first law of thermodynamics for closed systems. In contrast, open systems investigate the balance laws of mass and energy. This reasoning requirement prevents incorrect balance laws from being constructed based on three qualitative requirements. 
\begin{enumerate}
    \item Identification of orphan flows; output flows of conservable entities, i.e. mass (M) and energy (E), are not balanced by a reasonable input flow. 
    \item Identification of barren flows; input flows of conservable entities M and E are not balanced by a reasonable output flow. 
    \item Identification of changes in the state of a material when there is no additional or removal of energy. 
\end{enumerate}
Inspection I is performed on the BDD and the IBD to perform these checks, where all the blocks in the BDD and properties in the IBD are inspected. For example, Fig. \ref{fig:BL1-Hairdryer} displays an imbalance as the output material flow from intake grill's input port is sent to OUT\_E port, which expects a flow type of energy. Additionally, Fig. \ref{fig:BL2-Hairdryer} displays a barren flow as there is an input flow of type E but no output flow of type E. 
% The balance of flows must be maintained across both ends of an association. Here, the port OUT\_M" should not be connected to an "OUT\_E".
% The balance of flows must be maintained between the inputs and outputs of a block. Here,  the port "IN\_E" indicates energy is coming in, but there's no port to suggest energy is coming out.
\subsection*{Inspection II: Incomplete Topology}
In functional modeling, nodes are functions performed by different components within the system, and the edges are flows through the system \cite{sen2013physics}. However, for structural modeling, nodes (Properties) are system components, and flows (Ports and Associations) are the relationships between different system components.  Inspection II, also curated from Sen et al. \cite{sen2013physics}, demonstrates how these flows must be verified to exhibit complete topology. Validating that these nodes' rules for carrier relations are conserved. Flows must be provided by a function or supplied by the environment and must be used by a function or released to the environment. The two requirements this inspection is governed by is the identification of dangling tails and heads. 
\begin{enumerate}
    \item Identification of dangling tail; detects flows whose tail is not connected to a node. 
    \item Identification of dangling head; detects flows whose head is not connected to a node.
\end{enumerate}

\begin{figure} [h!tbp]
    \centering
    \includegraphics[width=\linewidth]{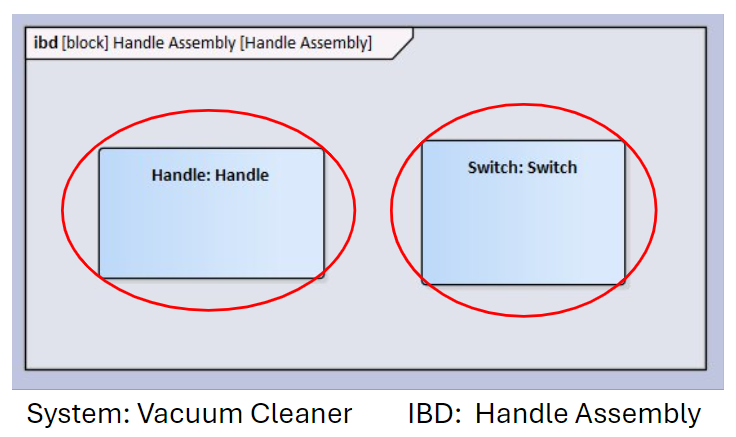}
    \caption{In handle assembly IBD, the handle and switch properties both have no flows}
    \label{fig:IT1-VacuumCleaner}
\end{figure}

\begin{figure} [h!tbp] 
    \centering
    \includegraphics[width=\linewidth]{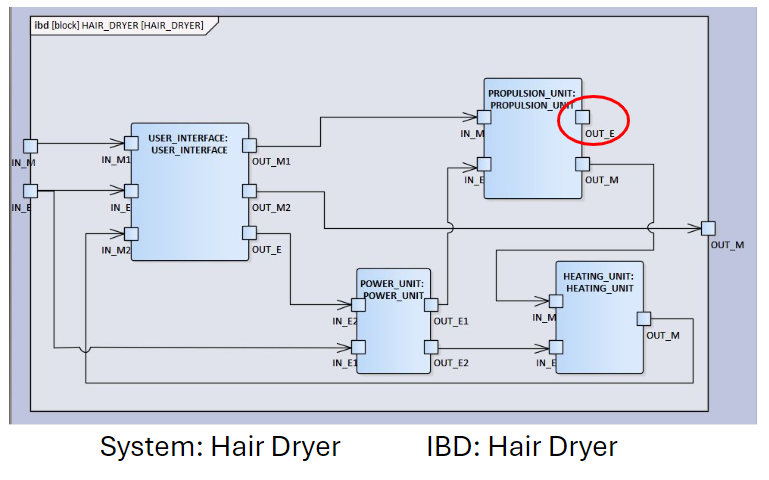}
    \caption{In hair dryer IBD, the propulsion unit has a dangling head as no association shows an energy flow being outputted from the propulsion unit.}
    \label{fig:IT2-Hairdryer}
\end{figure}

Inspection II is performed on IBDs, where each property, port, and association are inspected. For example, Fig. \ref{fig:IT2-Hairdryer} displays a dangling head as the port OUT\_E has no association, showing energy outputting out of the propulsion unit. On the other hand, Inspection II also captures structural dangling nodes; as seen in Fig. \ref{fig:IT1-VacuumCleaner}, it displays a structural dangling node as in the Handle Assembly IBD for the Vacuum Cleaner; both the handle and switch properties have no edges (ports or associations). 

% There should be no properties or blocks without ports because all components of the diagram should be related to flows. Here, the discharge filter and discharge grill have no ports so they are incomplete.
% The incomplete topology validation checks dangling tails and heads on properties. Here, the port "OUT\_E" is incomplete because it is not connected to any association.
\subsection*{Inspection III: Inferred Balance and Dangling Node based on the Knowledge Base}
Inspection III begins with evaluating if the nodes representing functions are not dangling. This investigation once again follows the Sen et al. criterion that defines a dangling node check as detecting nodes that do not input or output any flow with respect to the function and environment \cite{sen2013physics}. Secondly, this inspection validates the model to provide helpful feedback about how the model is balanced. Therefore, this check also infers qualitative balance between flow sets (input to and output) for each function and the model as a whole. Thus, this work verifies functional and flow integrity when a single function is decomposed into a graph or vice-versa \cite{ganeriwala2023functional}. A system functional knowledge base (KB) must be included to assist with performing this inspection. The KB curated is based on the Hirtz et al. functional basis set \cite{Hirtz2002} that consists of 53 function verbs and 45 flow nouns by systematically dissecting hundreds of products and naming their functions and flows. This KB is reconciled with 18 functions typically used to describe the functions of a system. Each verb is associated with the number of ports expected, the type of flows, and their topology.
\begin{figure} [h!tbp]
    \centering
    \includegraphics[width=\linewidth]{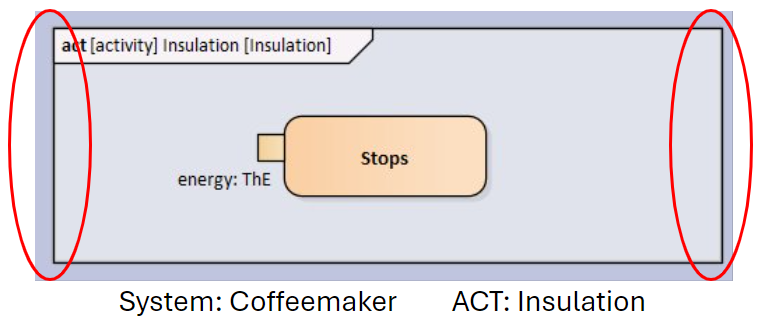}
    \caption{In the Insulation ACT, there are no input and/or output activity parameters in the diagram, indicating no input and output flows.}
    \label{fig:DN-Coffeemaker}
\end{figure}
% Each ACT should have at least one activity parameter. Here, the insulation ACT has no parameters so there are no flows coming in or going out which makes it a dangling node.
\begin{figure} [h!tbp]
    \centering
    \includegraphics[width=\linewidth]{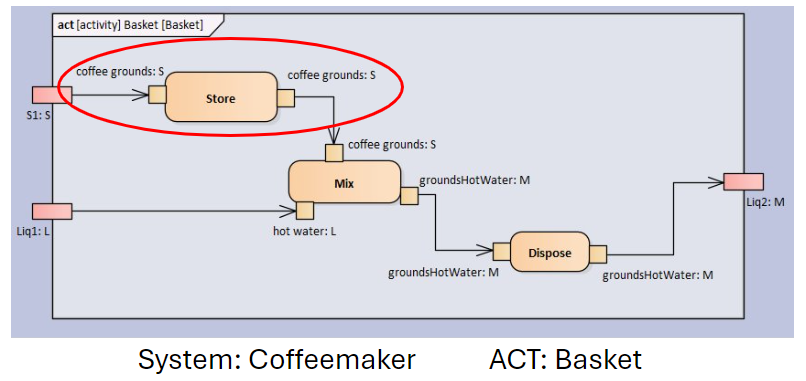}
    \caption{In the basket ACT, the store action displays both an input and output action, which violates the system functional KB}
    \label{fig:IB-Coffeemaker}
\end{figure}
% The number and type of inputs and outputs on each action block should correspond to the function definition specified in the knowledge base. Here, the "store" function should have no output, but instead, there are coffee grounds coming out.

Inspection III is performed on ACTs, where all the actions, transitions, action pins, and activity parameters are inspected. For example, Fig. \ref{fig:DN-Coffeemaker} represents a dangling node as there is an action that is expecting input from another function/environment; however, none is provided. Conversely, Fig. \ref{fig:IB-Coffeemaker} shows an inferred balance violation as for the store action, coffee grounds are being stored; however, there is an action pin and transition that shows the coffee grounds being supplied to the mix action. Therefore, according to system functional KB, the functional verb \emph{store} is expected to have only input and no output; hence, this diagram is flagged as an inferred balance.

\begin{figure*}[htpb!]
    \centering
    \includegraphics[width=\linewidth]{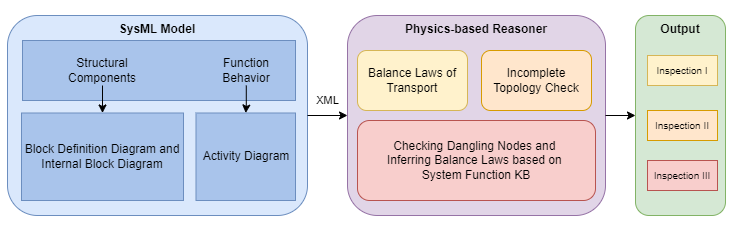}
    \caption{Automated Reasoner Framework}
    \label{fig:methodology}
\end{figure*}

\section{Automated Physics-Based Reasoner}
\label{sec:autoReason}
% Summer
The automated reasoner extracts structural and functional knowledge from the model and uses it to perform each of the previously discussed physics-based inspections. Prior to running the reasoner, the input XML file must be condensed to remove unnecessary information which was accomplished by using previous work by Chambers et al. \cite{chambers2023towards}. Then, components of the BDDs, IBDs, and ACTs are analyzed in the context of these validations. If an error is detected, it is stored corresponding to a particular component and displayed at the end of the program without halting. Note that the final algorithm (Function Inferred Balance) may only be performed on the model if ACTs are available. Otherwise, the program is limited to performing a structural analysis based on the components’ relationships and flows. The framework is as given in Figure \ref{fig:methodology}. The reasoner implementation and test case XML files can be found on our project repository \footnote{https://github.com/SummerMueller/Functional-Reasoning.git}.

\subsection{Extract Knowledge Algorithm}
The program implements the DOMParser interface to extract information about the structure and function of the model. The reasoner first parses through the condensed XML file to collect and store each element in separate lists based on whether the current element is a block, property, port, association, transition, activity parameter, action, and action pin. Each element is associated with its name and a unique identifier called XMI.ID in the XML representation. Depending on the type of element, all other relevant fields included in its nested tags are stored as compiled in Table \ref{removetaglist}. For instance, associations are stored with the XMI.ID of their source and destination ports along with its association name and XMI.ID. Likewise, properties and ports are also stored with the XMI.ID of their owner block. By representing the entire system in separate lists of its respective elements, desired relationships can be extracted as needed to perform each physics-based inspection.
\begin{center}
\begin{table}[h!]
\centering
\caption{Extracted attribute list of elements being stored.}
\begin{tabular}{||p{2.2cm}| p{5.8cm} ||} 
 \hline
 Element & Stored Attributes\\ [0.5ex] 
 \hline\hline
 Blocks & name, xmi.id, ports[] \\
 \hline
 Property & name, xmi.id, owner xmi.id\\
 \hline
 Port & name, xmi.id, owner xmi.id, reusesProperty\\
 \hline
 Association & name, xmi.id, sourceName, destinationName, source xmi.id, destination xmi.id\\
 \hline
 Action & name. xmi.id, owner xmi.id, pins[]\\
 \hline
 Action Pin & name, xmi.id, flowType, owner xmi.id\\
 \hline
 Activity Parameter & name, xmi.id, flowType, owner xmi.id\\
 \hline
 Transition & name, xmi.id, sourceName, targetName, source xmi.id, target xmi.id, sourceElementType, targetElementType\\
 \hline
 Activity Diagram & name, xmi.id, owner xmi.id, actions[]\\
 \hline
 Internal Diagram & name, xmi.id, owner xmi.id, element xmi.ids[]\\
  \hline
  % Class & isSpecification, ea\_ntype, isActive, version, date\_created, date\_modified, gentype, tagged, phase, author, complexity, product\_name, status, tpos, ea\_localid, ea\_eleType, style, \$ea\_xref\_property, visibility, isRoot, isLeaf, isAbstract \\ 
 % \hline
 % Association & style, linemode, linecolor, linewidth, seqno, headStyle, lineStyle, ea\_localid, ea\_sourceID, ea\_targetID, virtualInheritance, containment, sourcestyle, ea\_end, visibility, isRoot, isLeaf, isAbstract, aggregation, isOrdered, targetScope, changeable, isNavigable  \\
 % \hline
 % Collaboration & isAbstract, isSpecification, ea\_ntype, version, isActive, date\_created, date\_modified, gentype, tagged, phase, author, complexity, status, tpos, ea\_localid, ea\_eleType, style, \$ea\_xref\_property, visibility \\
 % \hline
 % Diagram & version, author, created\_date, modified\_date, type, ea\_localid, matrixitems, swimlanes, EAStyle\\
 % \hline
\end{tabular}
\label{removetaglist}
\end{table}
\end{center}
\subsection{Incomplete Topology Algorithm}
The reasoner inspects that no incomplete topology exists in the form of dangling tails and heads for properties inside an IBD. In other words, all ports of a property within an IBD should be connected by an association. In SysML, each directed association has a source and destination where material or energy flows from the source to the destination. These fields are stored with each association during the extractKnowledge function. In XML, to distinguish a port as input or output, the list of associations must be iterated over to identify the port as a source or destination. If not, the topology check is violated, and the reasoner flags the port. During each iteration, the port is sorted into one of two separate lists, destination (input) ports and source (output) ports to be accessed for other inspections.

\subsection{Balance Laws Algorithm}
To inspect the flow integrity of a system, it must be proven that the specific type of flow remains balanced across the source/destination ports of each association while the general flow types are conserved across the input/output ports of each block. The flow integrity of each association was verified by retrieving the names of its source and destination ports to determine their specific type (e.g., S = solid, Liq = liquid, ME = mechanical energy, and EE = electrical energy). If the source and destination ports’ types are not equal, the program raises an error at that association. For the second type of flow integrity validation, each block's input and output ports in the BDD are retrieved. Each flow type is then categorized more generally as either a material or energy. The program tracks and compares the number of input material flows versus output material flows, and if they do not match, then the program raises an error on that block. Similarly, the same procedure is repeated for the number of energy flows, and an error is raised if the final total per type is inconsistent. At this point, if a flow integrity error is labeled on a particular block that corresponds to an ACT, the program refers to the functional reasoning algorithm for clarification.

\subsection{Dangling Node Algorithm}
The dangling node validation ensures that each node contributes to the model's functionality. Therefore, at least one flow should come into or go out of every node. To ensure each node has a flow, the reasoner iterates over every ACT and checks that each has at least one activity parameter. Because activity parameters indicate flows, if there exists an ACT without any activity parameters, then the reasoner detects that ACT as a dangling node. So, the reasoner will raise a dangling node error corresponding to that ACT.
\begin{table*}[h!tbp]
  \caption{Displaying the number and type of errors for each use case}
  \label{errors-reasoner}
  \centering
    \begin{tabular}{lcccc}
    \toprule
 % & & \multicolumn{2}{c}{\textbf{5-way Acc.}} & \multicolumn{2}{c}{\textbf{20-way Acc.}} \\ 
 \textbf{Use Case} & \textbf{Number of Errors} &  \textbf{Type of Error} & \textbf{Error Found in Diagram} & \textbf{Size (LOC)} \\ 
 \midrule
 \textbf{Hair Dryer}  & 5 & I1-Balance Law II & BDD & 2147\\ 
                      & 1 & I2-Incomplete Topology II & IBD &
                      \\\\
        % \midrule
       \textbf{Wired Speaker} & 3 & I1-Balance Law II & BDD& 2057\\\\
        % \hline
        % \midrule
        \textbf{Coffeemaker} & 1 & I2-Incomplete Topology I & BDD & 5343\\
                             & 3 & I2-Incomplete Topology II & IBD, ACT\\
                             & 1 & I3-Dangling Node \\
                             & 4 & I3-Unknown Function & ACT\\
                             & 7 & I3-Inferred Balance \\\\
        % \hline
        % \midrule
        \textbf{Vacuum Cleaner} & 32 & I2-Incomplete Topology II & BDD, IBD& 1245\\\\
    \bottomrule
\end{tabular}
\end{table*}
\subsection{Function Inferred Balance Algorithm}
Each ACT contains at least one action block illustrating a function(s) being performed. Input and output flow types are stored as action pins similar to the ports on properties within IBDs. The number and type of flows on an action block are analyzed separately in terms of input and output action pins. Each action has unique specifications for the number and types of input/output flows; therefore, each action block is assessed according to the rules outlined in the functional knowledge base. If the number or type does not align with its defined specifications, the program raises an error on that action. However, if all actions within an ACT have been validated, then the corresponding block or node will automatically pass the balance laws and flow integrity inspections.

\section{Validation Case Study}
\label{sec:valid}
% Parth, Candice and Summer
\subsection{Use Cases}
For this experiment, four use cases are evaluated using the automated reasoner described in section \ref{sec:autoReason}. The four use cases investigated are an electric coffeemaker, a hair dryer, a vacuum cleaner, and a wired speaker. These four use cases are all within the electromechanical domain and manually modeled using the Enterprise Architect software \cite{enterprise}. 
\subsubsection*{Coffeemaker}
The electric coffeemaker is modeled based on the functional requirements adapted from the AADL representation by Chauhan et al. \cite{chauhan2022toward}. The system comprises an interconnected cooking unit and a storage unit modeled using a BDD and nine IBDs. The cooking unit includes a hydraulic unit, a heating unit, and a brewing unit. The interconnection of the cooking unit brews and transports the brewed coffee into the storage unit. In addition, to represent the functional requirements of the coffeemaker, these functions are constructed using 16 ACTs. 

\subsubsection*{Hair Dryer}
The hair dryer is modeled based on structural requirements; the system comprises a user interface, a heating unit, a propulsion unit, and a power unit modeled using a BDD and four IBDs. These units work together to pull in air from the environment to produce the hot air dispensed from the hair dryer. 
\begin{figure}
    \centering
    \includegraphics[width=\linewidth]{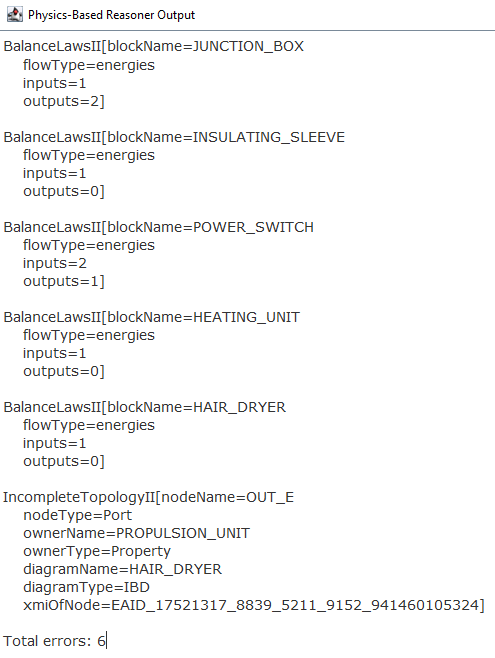}
    \caption{GUI displaying the hair dryer system inspection errors}
    \label{fig:reasoner}
\end{figure}
\subsubsection*{Vacuum Cleaner}
The vacuum cleaner is also modeled based on its structural requirements; the system contains an interconnected filter assembly, nozzle assembly, and drive assembly modeled using a BDD and 11 IBDs. These components work coherently to pull and store debris from its surroundings.

\subsubsection*{Wired Speaker}
Similarly to the hair dryer and vacuum cleaner, the wired speaker is modeled based on structural requirements. This system contains an interconnected enclosure and speaker terminal modeled using a BDD and six IBDs. The enclosure includes a midrange driver, tweeter driver, woofer driver, and a crossover network. These components demonstrate that electrical energy from the environment produces sound based on the input in the speaker terminal.

\subsection{Systematic Generation of Test Cases}
The errors within each system model were intentionally designed and embedded systematically during the development of the inspection process to ensure comprehensive evaluation by the automated reasoner. This approach aimed to test the robustness and reliability of the reasoner in detecting various types of errors. Each error was crafted to simulate real-world issues during system modeling, such as missing component connections, improper interactions between elements, incorrect allocation of functional requirements, and violations of standard modeling practices. By embedding these specific types of errors into the models, it was ensured that the automated inspection tool would be challenged to identify discrepancies that could occur in actual modeling scenarios. 

The test case generation followed a structured methodology; each use case was evaluated against a predefined set of inspection rules based on the physics-based principles and functional requirements outlined in the knowledge base (KB). The inspection process was configured to flag inconsistencies, missing relationships, or any deviation from the expected standards encoded in the system. This systematic approach not only validated the automated reasoner’s effectiveness in identifying these errors but also highlighted areas where enhancements could be made, such as expanding the lexicon of recognized functional verbs and improving its contextual understanding. By intentionally designing the errors, a controlled environment for assessing the reasoner’s capabilities and limitations is provided, facilitating targeted improvements to the tool and the underlying methodology.  
\subsection{Results}
Upon our investigation of the four use cases, the automated physics-based reasoner flagged inspection errors on each system. Figure \ref{fig:reasoner} displays the reasoner's graphical user interface (GUI) that displays all the six errors found in the hair dryer system modeled in SysML. Table \ref{errors-reasoner} shows that each system has its unique inspection errors; the coffeemaker, the largest of the four use cases, is modeled using BDD, IBDs, and ACT and has the most inspection errors. Because the coffeemaker is the only use case to represent function, it can portray inspection III, which evaluates functional requirements along the inspected physics-based rules. However, this reasoner limitation lies in its inability to perform spelling checks or synonym conversion. Therefore, this restricts the user to adhere only to the 18 functional verbs stored in our KB. If other functional verbs are used, this will be flagged as an unknown function error.  

\section{Conclusion}
\label{sec:conclusion}
In conclusion, this paper has demonstrated an automated approach to performing physics-based reasoning on systems modeled in SysML. This reasoner is built on physics-based principles following the first law of thermodynamics, incomplete topology, and inferred balances. In addition to these physics-based principles, a system functional knowledge base was also incorporated to further assist with flagging components that did not adhere to the inspections. The reasoner was evaluated on four use case scenarios modeled manually using the Enterprise Architect software. These systems were extracted to their XML format and input into the reasoner. Our results have shown that the reasoner accurately flags and identifies elements and their corresponding inspection rule for structural and functional diagrams for all test cases. In the future, the system's functional KB will be expanded to include synonyms for the existing functional verbs. In addition, this automated reasoner will be integrated to validate SysML models generated by an automated framework from English specification text.   
% Candice
\bibliographystyle{unsrt}
\bibliography{refs}

\begin{thebibliography}{10}
\bibitem{SysML}
Sysml faq: What is the systems modeling language (sysml)?
\end{thebibliography}
\end{document}